\documentclass[aps,pra,showpacs,superscriptaddress,preprint]{revtex4-1}
\usepackage{graphicx}
\usepackage{amsmath,bm}
\usepackage{xcolor,graphicx}

\usepackage{bm,,amsfonts}

\begin{document}
	
\title[Non-Markovianity in the optimal control]{Non-Markovianity in the optimal control of an open quantum system described by hierarchical equations of motion}

\author{E. Mangaud}
\affiliation{Laboratoire Collisions Agr\'egats R\'eactivi\'e (IRSAMC), Universi\'e Toulouse III Paul Sabatier, UMR 5589, F-31062 Toulouse Cedex 09, France}
	
\author{R. Puthumpally-Joseph}
\affiliation{Laboratoire Interdisciplinaire Carnot de Bourgogne (ICB), UMR 6303 CNRS-Universit\'e Bourgogne Franche Comt\'e, 9 Av. A. Savary, BP 47 870, F-21078 Dijon cedex, France}
\affiliation{Institut des Sciences Mol\'eculaires d'Orsay (ISMO) UMR CNRS 8214 ,Universit\'e Paris Saclay, Univ. Paris Sud,  F-91405 Orsay, France}

\author{D. Sugny}
\affiliation{Laboratoire Interdisciplinaire Carnot de Bourgogne (ICB), UMR 6303 CNRS-Universit\'e Bourgogne Franche Comt\'e, 9 Av. A. Savary, BP 47 870, F-21078 Dijon cedex, France}
\affiliation{Institute for Advanced Study, Technische Universit\"at M\"unchen, Lichtenbergstrasse 2 a, D-85748 Garching, Germany}

\author{C. Meier}
\affiliation{Laboratoire Collisions Agr\'egats R\'eactivi\'e (IRSAMC), Universi\'e Toulouse III Paul Sabatier, UMR 5589, F-31062 Toulouse Cedex 09, France}

\author{O. Atabek}
\affiliation{Institut des Sciences Mol\'eculaires d'Orsay (ISMO) UMR CNRS 8214, Universit\'e Paris Saclay, Univ. Paris Sud,  F-91405 Orsay, France}

\author{M. Desouter-Lecomte}
\affiliation{Laboratorie Chimie Physique (LCP)-CNRS,Universit\'e Paris Saclay, Univ. Paris Sud,  F-91405 Orsay, France}
\affiliation{D\'epartement de Chimie, Universit\'e de Li\`ege, Sart Tilman, B6, B-4000 Li\`ege, Belgium}

\date{\today}
	
\begin{abstract}
	Optimal control theory is implemented with fully converged hierarchical equations of motion (HEOM) describing the time evolution of an open system density matrix strongly coupled to the bath in a spin-boson model. The populations of the two-level sub-system are taken as control objectives; namely, their revivals or exchange when switching off the field. We, in parallel, analyze how the optimal electric field consequently modifies the information back flow from the environment through different non-Markovian witnesses. Although the control field has a dipole interaction with the central sub-system only, its indirect influence on the bath collective mode dynamics is probed through HEOM auxiliary matrices, revealing a strong correlation between control and dissipation during a non-Markovian process.
	A heterojunction is taken as an illustrative example for modeling in a realistic way the two-level sub-system parameters and its spectral density function leading to a non-perturbative strong coupling regime with the bath. 
	Although, due to strong system-bath couplings, control performances remain rather modest, the most important result is a noticeable increase of the non-Markovian bath response induced by the optimally driven processes.  
		
	\end{abstract}
	
	\maketitle



	\maketitle

	\section{Introduction}
	Open quantum systems (OQS) are ubiquitous in physics and chemistry and have many uses from setting quantum technology in condensed phase to exploring long-lived coherence in biological systems \cite{Weiss,Breuer,Vega,Rivas,MayKuhn, Chin}. They consist in selecting a given partitioning into a central quantum system and a statistical surrounding bath. The reduced system dynamics is non-unitary and can be called Markovian or non-Markovian according to the importance of memory effects \cite{Breuer}. The comparison of system and bath typical timescales is a relevant qualitative measure to separate both situations : if the timescale characterizing the bath is shorter than the one of the system, dynamics can be said Markovian, non-Markovian if not. For a two-level system, this characteristic time is the Rabi period whereas the bath dynamics can be estimated from the time decay of the two-time correlation function of the system bath coupling related to the Fourier transform of the bath spectral density. A nearly delta correlated bath leads to a Markovian behavior usually described by Lindblad \cite{Lindblad} or Redfield \cite{Redfield, Breuer, MayKuhn} approaches involving unidirectional relaxation. Non-Markovianity is described by strong quantum memory effects leading to temporary information back flow from the environment to the system. Several measures of non-Markovianity have been proposed and compared recently in the literature \cite{Vega, Rivas}. Among them one can mention the distinguishability of quantum states estimated by their trace distance that can transitively decrease during the relaxation, as opposite to a Markovian evolution in which it continuously increases \cite{BreuerLP, BreuerLPV}. Other non-Markovianity signatures refer to a re-amplification of the volume of accessible states during the decay process \cite{Paternostro}, the detection of a negative canonical decay rate \cite{Anderson, Anderson2}, or a non-monotonous time evolution of the system von Neumann entropy \cite{Haseli}. Even more importantly, the role of transitory information back flow in externally controlled dynamics remains an open issue and an active research area \cite{Potz,Potz2,Mancal,Saalfranck,Tremblay,Lidar,Cui,Brumer,Tai,Koch,Poggi,Glaser,koch2,Addis,molphys,pra,Sugny}. 
	
	\noindent
	Our main purpose is to take advantage of the back flow of information from the surrounding bath, characterizing non-Markovianity, to enforce the control of the central system physical observables, protecting them against decoherence. At that respect, the present paper is a second one of a series of three \cite{molphys, pra} where an optimal control scheme is worked out, still acting on the central system alone, aiming at some protection against decoherence (population revivals, or robust and efficient transfers) and subsequently examine its consequences in terms of the bath non-Markovian response. More precisely, we analyze non-Markovianity during an ultra-short field pulse optimized by quantum control \cite{Tannor, Shapiro, Kosloff} in a spin-boson (SB) model \cite{Weiss, Breuer, Leggett} where the active sub-system strongly interacts with the bath. The controlled dynamics ends before the complete decay of the volume of accessible states in the Bloch sphere \cite{Paternostro}, i.e. before the decay of the bath correlation function which means before quantum memory (or non-Markovian) effects are expected to vanish. The control is also shorter than the full relaxation time of the state populations towards equilibrium. The interaction of the two-level system with the bath is described by the standard spin boson Hamiltonian (SB) which can used in many different situations ranging from qubit in quantum dots to exciton or charge transfer. In the present work, it is built and calibrated to simulate a charge transfer between donor and acceptor electronic states in a heterojunction \cite{Tamura, Tamura2}. The model addresses ultra-short control of electronic dynamics in a complex system strongly coupled to the nuclear vibrational motion \cite{MayKuhn}. Similar coherent control of excitation energy transfers in photosynthetic systems has already been investigated, but in weak coupling regimes, referring to Markovian approaches \cite{Plenio, Whaley}. Here we analyze a non-perturbative situation, described through hierarchical equations of motion (HEOM) \cite{Kubo,Ishizaki,Tanimura,Xu,Schulten,Aspuru}. We focus on early dynamics and we investigate the extent to which optimal control field enhances non-Markovianity during control. The canonical decoherence rates and the von Neumann entropy are taken as signatures of non-Markovianity. In a recent work, the enhancement of non-Markovianity during laser driven dynamics has been studied with simple periodic fields in a SB model with a smooth Lorentzian spectral density \cite{Poggi}. This example shows an enhancement of non-Markovianity signatures but for weak coupling only. On the contrary, in the present work,  we obtain  non Markovian behaviours even in the strongly coupled case.  
	
	\noindent
	Optimal control theory (OCT) is implemented here together with the HEOM method. The Rabitz monotonous algorithm in Liouville space we are referring to \cite{Rabitz, Ohtsuki,Turinici, Yan}, requires the forward and the backward propagations of the master equation. The memory kernel occurring in a time non-local master equation with a final condition has been discussed in different works. It has been implemented at second order level keeping the memory kernel \cite{Ohtsuki,Manz,Ndong} and by the auxiliary matrix method leading to time local coupled equations \cite{Yan, Chenel}. We generalize here this methodology with HEOM equations at higher order. The HEOM master equation can be rewritten as a time dependent Lindblad superoperator with time dependent canonical rates to get a witness of non-Markovianity \cite{Anderson,Anderson2}. This interesting Krauss decomposition \cite{Krauss, Alicki} has already been suggested to analyze the control in ref.\cite{Lidar}. In a first attempt, we do not impose any constraint on the field area so that the optimal field is not necessarily an optical one with zero area \cite{Brabec,Dion,Bandrauk,Atabek}. Such constraint could be added in a second step, but this issue would go beyond the scope of this paper. The electric field is assumed to have a dipole interaction with the central system only. However, since the memory kernel depends on the external field through the system Hamiltonian, this latter has an influence on the bath dynamics so that control and dissipation are strongly correlated. The modification of the bath dynamics is probed here from the HEOM formalism by analyzing the first moment of the bath collective mode \cite{Shi}. 
	
	\noindent
	The paper is organized as follows. Section II describes the SB model calibrated from data simulating a charge transfer in a heterojunction. The HEOM equations, the signatures of non-Markovianity and the optimal control theory in dissipative system are presented in Section III. Section IV gives the results for three ultra-short control cases, two for which the target is the initial state itself (a revival), and one for which the control enforces a transition between the two levels. Finally, some perspectives are presented in Section V.

	\section{The Model}\label{sec:model}
		The spin-boson model is a two-level quantum system linearly coupled to a bosonic bath of harmonic oscillators at thermal equilibrium. The Hamiltonian reads 
		\begin{equation}\label{Eq:Hamil}
		H(t) = {H_S}(t) + {H_B} + {H_{SB}}
		\end{equation}
		where ${H_S}(t)=\delta/2 \sigma_z + W \sigma_z - \bm{\mu} E (t) $, ${H_B} = \frac{1}{2} \sum\nolimits_k {\left( {p_k^2 + \omega _k^2q_k^2} \right)} $  in mass weighted coordinates and ${H_{SB}} = S\sum\nolimits_k {{c_k}{q_k}} $. Atomic units are used with $\hbar = 1$. The system operator is $S = {\sigma _z}  $ with $\sigma _i $ operators taken as Pauli matrices. The control field $E(t) $ only acts on the two-level system and is assumed to be linearly polarized. The $\bm{\mu}$ matrix is the matrix of the corresponding component of the dipole operator. In the context of a charge transfer between a donor and an acceptor in a heterojunction, ${H_S}(t=0)$ corresponds to the diabatic representation of the two electronic states for which the parameters are estimated at the equilibrium geometry. The  diabatic parameters $ \delta $ and $ W $ are taken from a model heterojunction between oligothiophene and fullerene \cite{Tamura, Tamura2}. The inter fragment distance is fixed to $  R = 3 $\AA \; leading to $ \delta  = 0.21 $ eV and  $ W = 0.13 $ eV. The corresponding Rabi period is 12.3 fs and the eigenenergy gap is 0.33 eV. The dipole matrices are not calibrated from ab initio calculations and different dipole models have been used to discuss the stability of the observed behaviors. In this electron transfer framework, the bath is formed by all the normal modes of the two fragments (here 264). The harmonic frequencies are assumed to be the same in both electronic states but the equilibrium geometries differ by a distance $ d_k $. Taking the origin of bath coordinates at a middle position between these equilibrium points, the vibronic coupling coefficients are ${c_k} = {\omega _k}^2{d_k}/2 $ . 
		
		The bath is fully characterized by the spectral density
		\begin{equation}\label{Eq:Jw-1}
		J\left( \omega  \right) = \frac{\pi }{2}\sum\limits_k^{} {\frac{{{c_k}^2}}{{{\omega _k}}}} \delta \left( {\omega  - {\omega _k}} \right)
		\end{equation}
		leading to the two-time correlation function 
		\begin{equation}\label{Eq:ct-1}
		C\left( {t - \tau } \right) = T{r_B}\left[ {B\left( t \right)B\left( \tau  \right)\rho _B^{eq}} \right] = \frac{1}{\pi }\int\limits_{ - \infty }^{ + \infty } {d\omega \frac{{J\left( \omega  \right){e^{i\omega \left( {t - \tau } \right)}}}}{{{e^{\beta \omega }} - 1}}}\,,
		\end{equation}
where $B\left( t \right) = \exp \left( {i{H_B}t} \right)B\exp \left( { - i{H_B}t} \right) $ is the bath operator $ B = \sum\nolimits_k {{c_k}{q_k}}$   in the Heisenberg representation. $\rho _B^{eq} = \exp \left( { - \beta{H_B}} \right)/T{r_B}\left[ {\exp \left( { - \beta {H_B}} \right)} \right] $   is the Boltzmann equilibrium density matrix of the bath and $\beta =1/k_BT$.  Spectral density and correlation functions (real, imaginary parts and modulus) of this heterojunction model are displayed in Fig.1. In this example, the Rabi period (12.3 fs) is smaller than the correlation time (25 fs) so that non-Markovian dynamics is expected.

		\begin{figure}[!ht]
			\centering
			\includegraphics{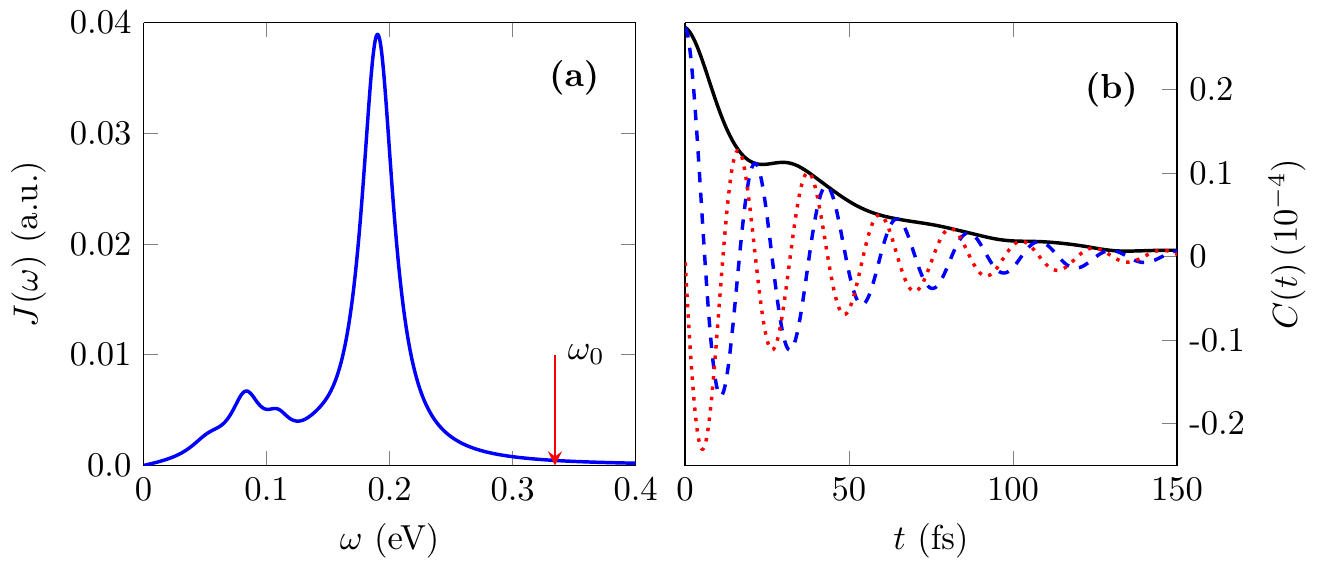}
			\caption{ Panel (a) : Spectral density of the spin-boson model. The red arrow marks the value of the system energy gap. Panel (b): Correlation function of the bath mode for $T$ = 298 K. The blue dashed curve is the real part and the red dotted curve is the imaginary part of the correlation function $ C(t) $. The black solid line is the absolute value of $ C(t) $.}
			\label{fig:jw_ct}
		\end{figure}
		\noindent
		As displayed in Fig.\ref{fig:jw_ct}, the spectral density $ J(\omega) $ is fitted by four four-pole functions
		
		\begin{equation}\label{Eq:jw-2}
		J\left( \omega  \right) = \sum\limits_{l = 1}^{{4}} {\frac{{{p_l}{\omega ^3}}}{\Lambda_{l,1}(\omega)\,\Lambda_{l,2}(\omega)}}\,
		\end{equation}
		where
		
		\begin{equation}\label{Eq:jw-2_1}
		\Lambda_{l,(1,2)}(\omega) = \left[ {{{\left( {\omega  + {\Omega _{l,(1,2)}}} \right)}^2} + \Gamma _{l,(1,2)}^2} \right]\left[ {{{\left( {\omega  - {\Omega _{l,(1,2)}}} \right)}^2} + \Gamma _{l,(1,2)}^2} \right]
		\end{equation}
		
		\noindent
		Cauchy's residue theorem is used to compute the integral of Eq.(\ref{Eq:ct-1}) with a contour closed in the upper half-plane enclosing $ 4n_l $ poles in $\left( {{\Omega _{l,1}},{\Gamma _{l,1}}} \right)$, $\left( { - \,{\Omega _{l,1}},{\Gamma _{l,1}}} \right)$, $\left( {{\Omega _{l,2}},{\Gamma _{l,2}}} \right)$, $\left( { - \,{\Omega _{l,2}},{\Gamma _{l,2}}} \right)$ and an infinity of poles on the imaginary axis $\left\{  \forall j \in \mathbb{N}^*/\left( {0,{\nu _j} = \frac{{2\pi }}{\beta }j} \right) \right\}$ called the Matsubara frequencies. In practice, the number of Matsubara terms is limited ensuring convergence for a given temperature. 
		
\section{Methods}\label{sec:methods}
\subsection{HEOM equations}
The system density matrix is the partial trace of the full density matrix $ \Xi (t) $ over the bath degrees of freedom $ \rho (t) = T{r_B}\left[ {\Xi (t)} \right] $. The initial condition is assumed to be factorized $ \Xi (t = 0) = \rho (t = 0){\rho _B^{eq}} $. The hierarchical equations of motion have been established from the path integral method \cite{Xu} or from the stochastic Liouville equation \cite{Kubo,Ishizaki,Tanimura}. The non-Markovian master equation
\begin{equation}\label{Eq:master}
\dot \rho (t) =  - iTr_{B} (\left[ {H,\Xi (t)} \right]) 
\end{equation}
is solved by a time local system of coupled equations among auxiliary matrices arranged in a hierarchical structure. 
\noindent
The algorithm requires a particular parametrization of the correlation function as a sum of $ n_{cor} $ exponential terms, written as:
\begin{equation}\label{Eq:ct-2}
C\left( {t - \tau } \right) = \sum\limits_{k = 1}^{{n_{cor}}} {{\alpha _k}{e^{i{\gamma _k}\left( {t - \tau } \right)}}}
\end{equation}
Analytical expressions for the $  \alpha_k $ and $ \gamma_k $ parameters can be derived when the spectral density is fitted by a sum of two-poles \cite{Meier} or four-pole Lorentzian functions leading to an Ohmic or super Ohmic behavior at low frequencies \cite{Mangaud}. The complex conjugate of the correlation function can be expressed by keeping the same coefficients $ \gamma_k $ in the exponential functions but using modified coefficients $ \tilde{\alpha}_k $ according to:
\begin{equation}\label{Eq:ct_*}
{C^*}\left( {t - \tau } \right) = \sum\limits_{k = 1}^{{n_{cor}}} {{{\tilde \alpha }_k}} {e^{i{\gamma _k}\left( {t - \tau } \right)}}
\end{equation}
$k$ being a collective index such that, $\tilde{\alpha}_{l,1}=\alpha^{*}_{l,2} $, ${\tilde \alpha _{l,2}} = \alpha _{l,1}^*$, ${\tilde \alpha _{l,3}} = \alpha _{l,4}^*$, $ {\tilde \alpha _{l,4}} = \alpha _{l,3}^* $ and $ {\tilde \alpha _{j,matsu}} = {\alpha _{j,matsu}} $, where $ \alpha_{l,m},\tilde \alpha_{l,m} $ with $ m =1,4 $ are related to the four poles of each Lorentzian $l$ \cite{Pomyalov}.

\noindent
The level $ L $ of the hierarchy corresponds to an order $ 2L $ in the perturbation expansion of the initial non-Markovian equation. Auxiliary matrices are labeled by a collective index ${\bf{n}} = \left\{ {{n_1}, \cdots ,{n_{{n_{cor}}}}} \right\}$ specifying the number of occupation of each artificial mode associated with one of $  n_{cor} $ decaying components. The system density matrix $\rho(t)$ has the index $ {\bf{n}} = \left\{ {0, \cdots ,0} \right\} $. The first level $ L=1 $ contains  $  n_{cor} $  auxiliary matrices with a single excitation only $ \sum\nolimits_k {{n_k}}  = 1 $. The HEOM coupled differential equations are given by  : 

\begin{eqnarray}\label{Eq:HEOM}
{\dot \rho _{\bf{n}}}(t) & = & - i\left[ {{H_S}(t),{\rho _{\bf{n}}}(t)} \right] + i\sum\limits_{k = 1}^{{n_{cor}}} {{n_k}{\gamma _k}{\rho _{\bf{n}}}\left( t \right)} \nonumber \\ 
& - & i\left[ {S,\sum\limits_{k = 1}^{{n_{cor}}} {{\rho _{{\bf{n}}_k^ + }}\left( t \right)} } \right]  - i\sum\limits_{k = 1}^{{n_{cor}}} {{n_k}\left( {{\alpha _k}S{\rho _{{\bf{n}}_k^ - }} - {{\tilde \alpha }_k}{\rho _{{\bf{n}}_k^ - }}S} \right)} 
\end{eqnarray}
with $ {\bf{n}}_k^ +  = \left\{ {{n_1}, \cdots ,{n_k} + 1, \ldots ,{n_{{n_{cor}}}}} \right\} $ and $ {\bf{n}}_k^ -  = \left\{ {{n_1}, \cdots ,{n_k} - 1, \ldots ,{n_{{n_{cor}}}}} \right\}$. Each matrix is coupled only to the superior and inferior levels in the hierarchy. The level of the hierarchy is chosen until convergence is reached for the system density matrix. 

\noindent
The HEOM formalism allows one to get insight into the correlated system-bath dynamics by probing the different moments $ {X^{(n)}}(t) = T{r_B}\left[ {{B^n}{\Xi}(t)} \right] $  of the collective mode $ B = \sum\nolimits_i {{c_i}} {q_i} $ \cite{Shi}. In particular, the expectation value of $B$ in each state is given by the diagonal elements of the $X^{(1)}(t)$ operator given by the sum of the first level auxiliary matrices
\begin{equation}\label{Eq:auxiliary}
{X^{(1)}}(t) =  - \sum\nolimits_{{\bf{n}}} {{\rho _{{\bf{n}}}}} (t)\,,
\end{equation}
where the sum runs over all index vectors $ {\bf{n}} = \left\{ {{n_1}, \cdots ,{n_{{n_{cor}}}}} \right\} $ with $ \sum\limits_l {{n_l}}  = 1 $. Recursive formula for higher orders can be found in ref. \cite{Shi}. This first moment already provides a signature of the induced correlated system-bath dynamics. As discussed in \cite{Shi}, the master equation can be recast to emphasize the role of $ {X^{(1)}}(t) $  in the system dynamics by writing 
\begin{equation}
 \dot \rho (t) =  - i\left[ {{{H}_S},\rho (t)} \right] + i\left[ {S,{X^{(1)}}(t)} \right].
\end{equation}

\subsection{Non-Markovian witnesses.} 
Signature of non-Markovianity is discussed here through the volume of the accessible states \cite{Paternostro} and through the canonical decoherence rates of a time-dependent Lindblad form  \cite{Anderson, Anderson2}. In the two-level case, the dynamical map $ \rho (t) = {\phi _t}\left[ {\rho (0)} \right] $ is first expressed in the basis set of the $ d^2 $  Hermitian operators (here $d = 2 $) formed by the identity $ {G_0} = {\bf{I}}/\sqrt{d}  $ and three operators $ G_m $ with $ m =1, 3 $ which are the Pauli matrices  $ \sigma_{x,y,z}/\sqrt{d} $. The equation then becomes
\begin{equation}\label{Eq:rho-1}
\rho (t) = \sum\nolimits_{k = 0}^{{d^2} - 1} {Tr\left( {{G_k}\rho (0)} \right)} {\phi _t}\left[ {{G_k}} \right]
\end{equation}
The volume of accessible states may be obtained from the matrix representation of the dynamical map in this basis set ${F_{m,n}}(t) = Tr\left( {{G_m}{\phi _t}\left[ {{G_n}} \right]} \right)$ by
\begin{equation}\label{Eq:Vol-1}
V(t) = \det \left( {\bf{F}} \right).
\end{equation}
This volume may also be expressed as a function of the decoherence canonical rates. The master equation is then recast in a canonical Lindblad form but with time dependent rates associated with time-dependent decay channels. Details can be found in refs \cite{Anderson, Anderson2}. The master equation is reformulated as 
\begin{equation}\label{Eq:rho-2-0}
\dot \rho (t) =  - i\left[ {{{H}_{S}},\rho (t)} \right] + \sum\limits_{j,k = 0}^{{d^2} - 1} {{a_{jk}}(t)}  {{G_j}\rho (t){G_k}} 
\end{equation}
In order to describe the decrease of the Bloch volume independently of the translation of its center, the contribution of the unity operator is separated by gathering terms containing coefficients $a_{j0}$. One then defines an operator $O = a_{00}/2d + \sum\limits_{i = 1}^{{d^2} - 1} (a_{i0}/d^{1/2})G_i$ and a corrected system Hamiltonian $ {H}_{S\,cor} = i \hbar (O - O^\dag)/2$. The relaxation operator then involves only the three operators associated with the Pauli matrices and the master equation takes the form :
\begin{equation}\label{Eq:rho-2}
\dot \rho (t) =  - i\left[ {{{H}_{S\,cor}},\rho (t)} \right] + \sum\limits_{j,k = 1}^{{d^2} - 1} {{D_{jk}}(t)} \left( {{G_j}\rho (t){G_k} - \frac{1}{2}\left\{ {{G_k}{G_j},\rho (t)} \right\}} \right)
\end{equation}
where $ D_{jk}(t) $  is the decoherence matrix. Its diagonalization provides the decoherence canonical rates $ g_k(t) $  and the decay channels $ C_k(t) $. Eq. (\ref{Eq:rho-2}) becomes 
\begin{eqnarray}\label{Eq:rho-3}
\dot \rho (t) =  -i\left[ {{{\hat H}_{Scor}},\rho (t)} \right]\nonumber\\
+ \sum\limits_{k = 1}^{{d^2} - 1} {{g_k}(t)} \Big{(} {2{C_k}(t)\rho (t)C_k^\dag (t) - \left\{ {C_k^\dag (t){C_k}(t),\rho (t)} \right\}} \Big{)}\,,
\end{eqnarray}
with $ {D_{ij}}(t) = \sum\nolimits_{k = 1}^{{d^2} - 1} {{U_{ik}}} (t){g_k}(t)U_{jk}^*(t) $ and $ {C_k}(t) = \sum\nolimits_{i = 1}^{{d^2} - 1} {{U_{ik}}} (t){G_i} $.

\noindent
It is worthwhile noting that the occurrence of negative canonical decoherence rates $ g_k(t) $  yields another characterization of non-Markovianity \cite{Anderson}. The rates are linked to the volume of accessible states through the relation 
\begin{equation}\label{Eq:Vol-2}
V(t) = V(0)\exp \left( { - d\int_0^t {\Gamma (s)ds} } \right)
\end{equation}
with
\begin{equation}\label{Eq:Partial_gamma}
\Gamma (t) = \sum\nolimits_{k = 1}^{{d^2} - 1} {{g_k}(t)} 
\end{equation}
The criterion based on the volume can be considered as an average measure since it depends of the sum of the rates only. Thus, it can be considered as a less stringent witness of non-Markovianity than a negative canonical decoherence rate $g_k(t)$.

\noindent
A possible numerical strategy to compute the decoherence matrix  $ D_{ij}(t) $ has been discussed in \cite{Anderson} and is given by 
\begin{equation}\label{Eq:DecohMat_1}
{D_{ij}}(t) = \sum\nolimits_{m = 0}^{{d^2} - 1} {Tr\left[ {{G_m}{G_i}{\Lambda _t}\left[ {{G_m}} \right]{G_j}} \right]} 
\end{equation}
with 
\begin{equation}\label{Eq:DecohMat_2}
{\Lambda _t}\left[ {{G_j}} \right] = \sum\nolimits_{k = 0}^{{d^2} - 1} {{{\dot \phi }_t}} \left[ {{G_k}} \right]F_{kj}^{ - 1}\,.
\end{equation}

\noindent
Besides the analysis of the decoherence canonical rates, we also compute the von Neumann entropy of the system that should vary monotonously in a Markovian evolution \cite{Haseli}
\begin{equation}\label{Eq:Entropy}
S(\rho (t)) =  - Tr\left[ {\rho (t){{\log_2 }}\rho (t)} \right] =  - \sum\nolimits_k {{\lambda _k}{{\log_2 }}} {\lambda _k}\,,
\end{equation}
where $ \lambda_k $ are the eigenvalues of the system density matrix.

\subsection{Optimal Control Theory}
We use optimal control theory in the Liouville space \cite{Rabitz, Ohtsuki, Yan} to optimize the field driven state-to-state transfer at the end of the pulse of total duration $ t_f $. The cost functional is built from a chosen performance index from state to state at time $t_f$, here $ Tr\left[ \rho^\dag({t_f})\rho _{target} \right] $ with a contraint on the field intensity and one assuring the respect of the mater equation at any time. The corresponding Lagrange multipliers are the scalar $ \alpha_0 $  and the density matrix $ \chi(t) $  respectively. We do not enforce here the constraint on the pulse area which is required for a purely optical field \cite{Atabek}. The optimal field is obtained from the system matrix density propagated by the master equation with initial condition $ \rho (t = 0) = {\rho _{ini}} $ and from the Lagrange multiplier propagated with a final condition $ \chi (t = {t_f}) = {\rho _{target}} $. 
The corresponding master equations with initial and final conditions take the form with 
$L \bullet=-i\left[{H_S}(t),\bullet\right] $
\begin{eqnarray}
\dot \rho (t) = L\rho (t) + \int_0^t {K(t,t')\rho (t')dt'} \label{Eq:Opt-1} \\
\dot \chi (t) = L\chi (t) - \int_t^{{t_f}} {{K^\dag }} (t,t')\chi (t')dt'\,.\label{Eq:Opt-2}
\end{eqnarray}

When the master equation is solved by the HEOM algorithm, the operational equations for the Lagrange multiplier can be derived by using Eqs.(\ref{Eq:ct-2}) and (\ref{Eq:ct_*}) 
\begin{eqnarray}\label{Eq:Opt-3}
{\dot \chi _{\bf{n}}}(t) &=& L{\rho _{\bf{n}}}(t) - i\sum\limits_{k = 1}^{{n_{cor}}} {{n_k}{\gamma _k}{\rho _{\bf{n}}}\left( t \right)}  \nonumber\\
&-& i\left[ {S,\sum\limits_{k = 1}^{{n_{cor}}} {{\rho _{{\bf{n}}_k^ + }}\left( t \right)} } \right]  + i\sum\limits_{k = 1}^{{n_{cor}}} {{n_k}\left( {{\alpha _k}{\rho _{{\bf{n}}_k^ - }}S - {{\tilde \alpha }_k}S{\rho _{{\bf{n}}_k^ - }}} \right)} \,.
\end{eqnarray}
In practice Eq.(\ref{Eq:Opt-3}) is solved backwards starting from $ {\chi _{\left\{ {0,0,..,0} \right\}}}(t = {t_f}) = {\rho _{target}} $ with all the auxiliary matrices set equal to zero. The field at iteration $k$ is obtained by $ {E ^{(k)}} = {E ^{(k - 1)}} + \Delta {E ^{(k)}} $, where $ \Delta {E ^{(k)}} $ is estimated by \cite{Rabitz}

\begin{equation}\label{Eq:Opt-Field}
\Delta E (t) = \frac{1}{{{\alpha _0}}} \Im m \left\{ {Tr}\left( {\rho (t)\chi (t)} \right)Tr\left( {\chi (t)\left[\bm{\mu},\rho (t) \right]}  \right)\right\}\,.
\end{equation}

\section{Results}\label{sec:Results}
HEOM equations are solved using a Cash-Karp adaptative stepsize Runge-Kutta algorithm with a small time step of 2 a.u. during which the field is assumed to be constant. Dynamics converges at level $L=6$ of the HEOM hierarchy, \emph{i.e.} at order 12 in perturbation theory which shows a strong system-bath coupling. In the above examples the dipole matrix is merely set equal to $ \bm{\mu}= \mu {\sigma _z} $ with $ \mu=1 $ a.u. Stability of the results has been verified for different non diagonal dipole matrices. The guess field is a sine square with maximum amplitude $ 10^{-3} $ a.u. The duration of the control is fixed to $ 20 $ fs, smaller than typical times for the complete field free decay of the Bloch sphere volume (Eq.\ref{Eq:Vol-1}). No constraint on the shape of the field is imposed by the OCT algorithm, except a penalty factor in such a way that the field amplitude does not exceed $ 10^{-2} $ a.u. ($3.51$ $10^{12}$ W/cm$^2)$.

\subsection{Field-controlled dynamics}
We consider three control objectives defined by the populations of the system. In the two first strategies that are denoted C1-1 and C2-2, the target is the revival of initial zero-order state, either state 1 or 2, at the end of the control. A third control  denoted C1-2, enforces the fast decay from state 1 to state 2 (a fast switch from 1 to 2). We compare the control without or with dissipation and analyze both the system and bath responses (memory effects) during the corresponding field-driven dynamics. 

\begin{figure}[!ht]
	\centering
	\includegraphics[width=0.95\textwidth]{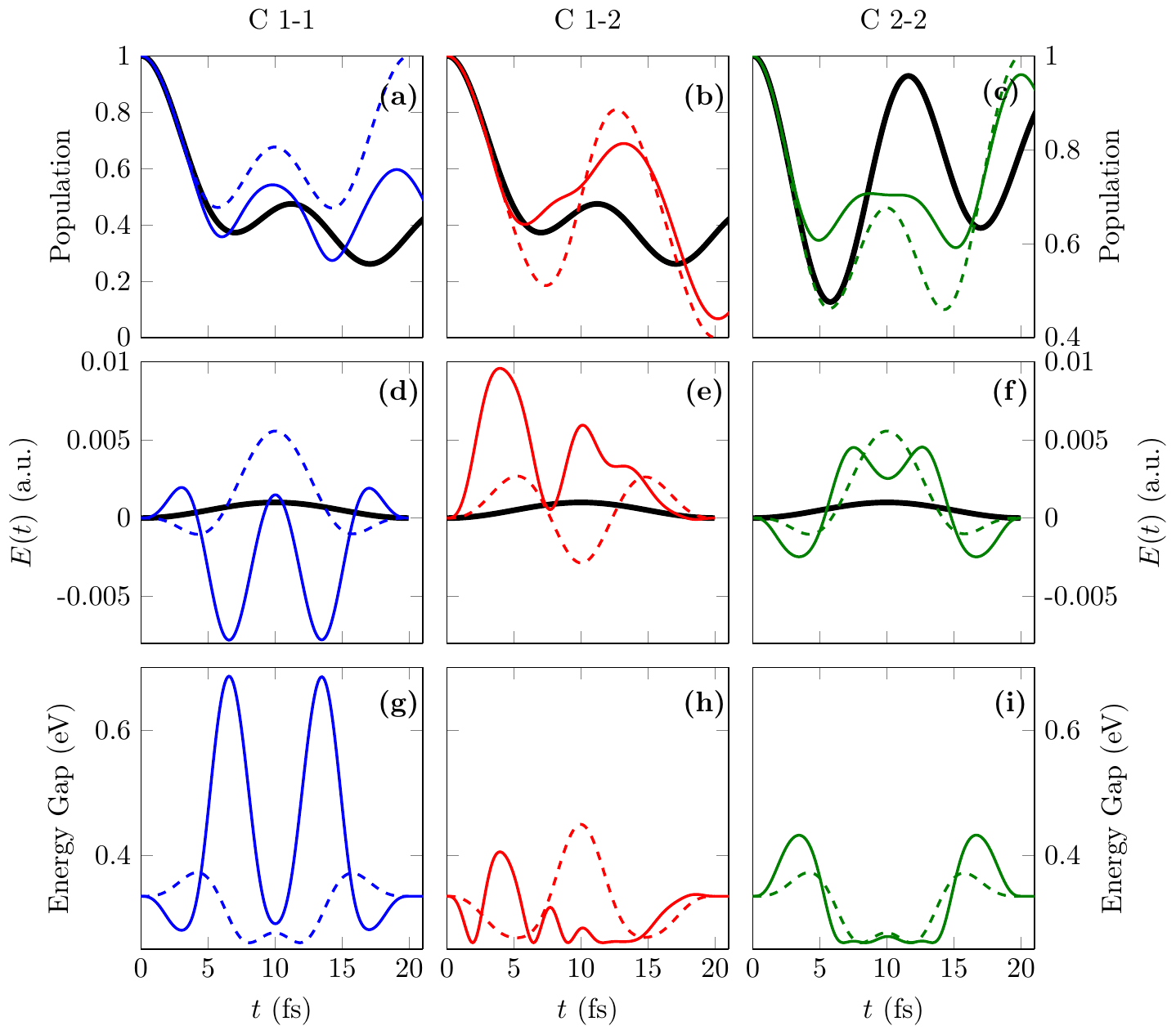}
	\caption{(Color online) The three columns correspond to the three control strategies C1-1, C1-2 and C2-2 respectively. Upper panels: Evolution of the population in the initial state (black full line: field-free dynamics, colored full lines: OCT with dissipation, dashed lines: OCT without dissipation).  Middle panels : Amplitude of the electric fields (black solid line : guess field, colored full lines: optimal field with dissipation, dashed lines: optimal field without dissipation). Lower panels : Fluctuations of the instantaneous system eigenenergy gaps $\omega_0(t)$ induced by the corresponding optimal fields (full lines: OCT with dissipation, dashed lines: OCT without dissipation)}
	\label{fig:pop}
\end{figure}

\noindent
The field free and field driven populations in the initial state during the three control strategies are shown in the upper panels (a), (b) and (c) of Fig.\ref{fig:pop}. The field free evolution (full black lines in Fig.\ref{fig:pop}) displays the expected damped Rabi oscillations of 12.3 fs. The dashed lines in panels (a) for C1-1,(b) for C1-2 or (c) for C2-2 are the populations driven by the optimal fields without dissipation. The objective is then reached easily with a performance index of 1. When the system is coupled to the bath, the populations are the full lines (blue for C1-1, red for C1-2 and green for C2-2). Panel (a) shows, for C1-1 strategy, at almost all times (except between 12 and 15 fs) a field enhanced protection of the population of the initial state 1 resulting in about $10\%$ of increase at the end of the control with respect to the field-free case. Similar final results are obtained for C2-2 illustrated panel (b) and C1-2 (panel (c)) but  their final results nearly matches  their dedicated target. In the isolated system, the only possible mechanism should be a modification of the oscillation periods, a decrease in the C1-1 or C2-2 scenario and an increase in the C1-2 case. This can be related to the transient variation of the energy gap induced by the control. In presence of dissipation, the variation of the gap acts both on the period and on the strength of the system-bath coupling. Panels (d), (e) and (f) in Fig.\ref{fig:pop} present the corresponding optimal fields in full lines for control with dissipation and in dots without dissipation. The profiles are very symmetrical for the two control strategies C1-1 and C2-2 (due to the same kind of optimal control with identical initial and final conditions), which is not the case for C1-2 with dissipation. The field-free energy gap is 0.33 eV and the optimal fields induce different Stark shifts in a range of about 0.25 to 0.65 eV, so that the instantaneous resonance frequency $\omega_0(t)$ moves with respect to the spectral density peaks, with its expected consequences on non-Markovianity \cite{molphys,pra}. The fluctuations of the eigenenergy gap of the system field-dressed Hamiltonian are shown in panels (g), (h) and (i) of Fig.\ref{fig:pop}. Convergence has been checked by inverting the sign of the initial field : this leads to nearly the same final shape of the optimal fields. The mechanism found by the control exploits transitory decrease of the energy gap leading to region where the coupling with the bath increases and transitory strong increase of the gap leading to a decrease of the bath coupling but probably an enhancement of non-Markovian effects.

\begin{figure}[!ht]
	\centering
	\includegraphics{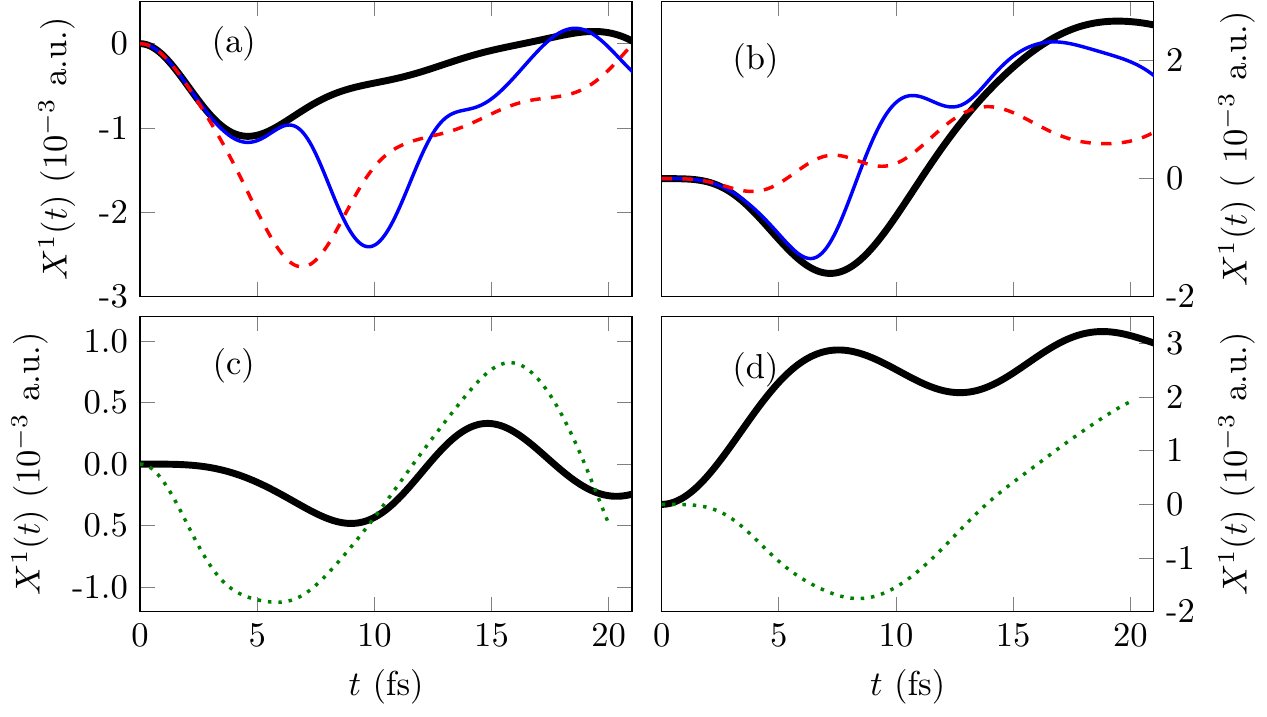}
	\caption{(Color online) Evolution of the diagonal elements of the $ X^{(1)}(t)$ operator giving the first moment of the bath collective mode. Left panels (a and c) for $X^{(1)}_{1,1}$ and right panels (b and d) for $X^{(1)}_{2,2}$. Upper panels, for initial state 1 : field free in full black lines, control C1-1 in full blue lines and C1-2 in dashed lines. Lower panels, for initial state 2 : field free in full black lines, control C2-2 in dotted lines. }
	\label{fig:moments}
\end{figure}

Obviously, control performances remain rather modest. This point can be explained both by limitations of the control parameters  (rather low field amplitudes and short pulse duration), and more importantly, by the way the strong system-bath interactions inherent to the specific molecular situation at hand interplays with the control. This can be numerically rationalized through the analysis of the first moment of the bath collective mode in each state given by the diagonal elements of the 2 $\times$ 2  $X^{(1)}$ matrix as depicted in Eq.(\ref{Eq:auxiliary}). 
Although the control field is explicitly introduced only to act on the system Hamiltonian, it affects the overall dynamics through the memory terms included in the right-hand-side of Eq.(\ref{Eq:HEOM}). Control fields indirectly modify the bath response leading to a strong correlation between control and dissipation. 
This is illustrated in Fig.\ref{fig:moments} which displays the first moment of the bath collective mode, in terms of the diagonal elements $X^{(1)}_{1,1}$ (left column) and $X^{(1)}_{2,2}$ (right column), starting either from initial state 1 (upper line) or 2 (lower line). The first observation is that bath oscillations roughly follow the field driven modifications of the Rabi period, with some amplitude and period variations. But marked differences are depicted according to the initial state.
For initial state 1, the short time dynamics (up to about 5 fs) is such that the field-controlled bath motions follow their field-free counterparts. Discrepancies from the field-free behaviors occur with opposite signs for $X^{(1)}_{1,1}$ and $X^{(1)}_{2,2}$, starting from the time when the gap is at its maximum value, i.e., close to 6 fs for C1-1 and 4 fs for C1-2 control strategies. Actually, when dealing with these two strategies, the gap is decreasing during the first femtoseconds, such that the system internal transition frequencies better match bath resonant phonons transitions. As a consequence, the amplitudes of collective modes oscillations are expected to increase.
For initial state 2 and the corresponding C2-2 control strategy, early Stark shifts have an opposite sign leading to increasing gaps, preventing bath resonant processes from occuring. Discrepancy from the field-free situation occurs at the very beginning of the control process. Such observations on the first moment $X^{(1)}$ can be considered as additional insight for a comprehensive rationalization of control strategies as they evolve in time. Actually, it turns out that control fields take advantage from two simultaneous mechanisms: (i) population transfer improved by modifying the Rabi frequency, through the Stark shift directly affecting the central system; (ii) dynamical decoupling effects, through indirect process in the bath, preventing overall decoherence. It is worthwhile noting that, we have previously reported similar mechanisms with single cycle or dc fields \cite{pra}. 
As a final remark, these mechanisms being dynamically mixed, a non-Markovian diagnostic cannot merely be inferred from their analysis. This motivates the need to resort to other non-Markovian witnesses as is done hereafter.  

\subsection{Non-Markovian signatures. }
\begin{figure}[!ht]
	\centering
	\includegraphics{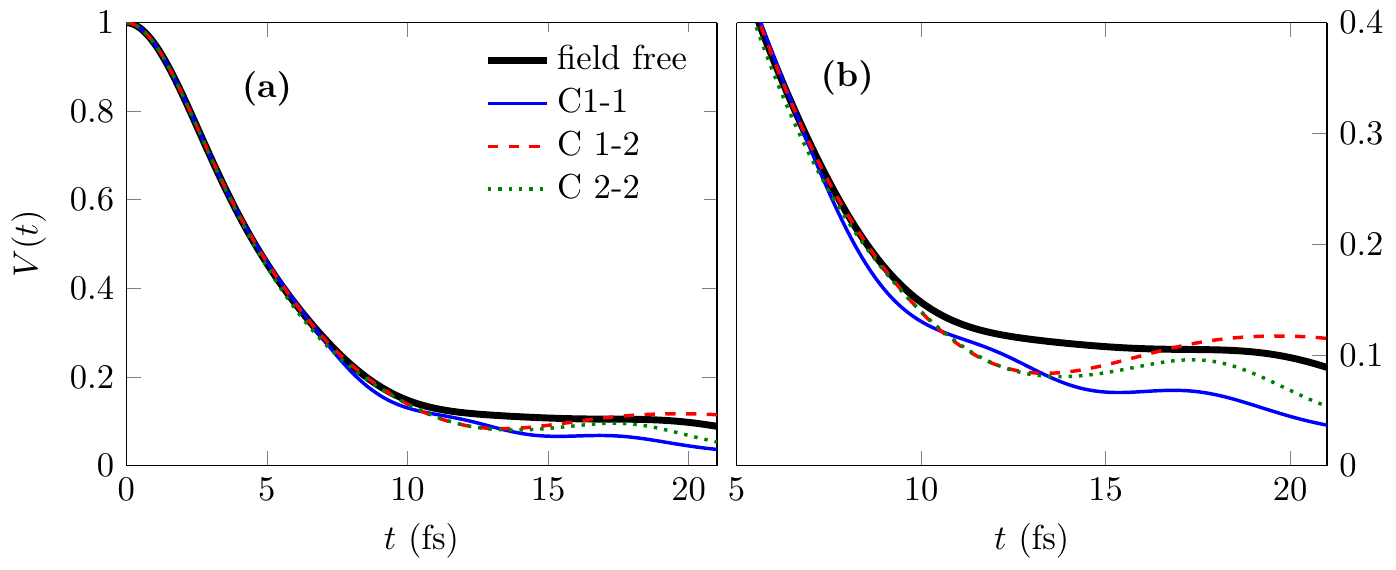}
	\caption{(Color online) Panel (a) volume of accessible states for field-free (solid black line) or field-controlled dynamics using strategies C1-1 (solid blue line), C2-2 (dotted green line) and C1-2 (dashed red line). Panel (b) is a zoom of (a) for times larger than 5 fs.}
	\label{fig:vol}
\end{figure}

\begin{figure}[!ht]
	\centering
	\includegraphics{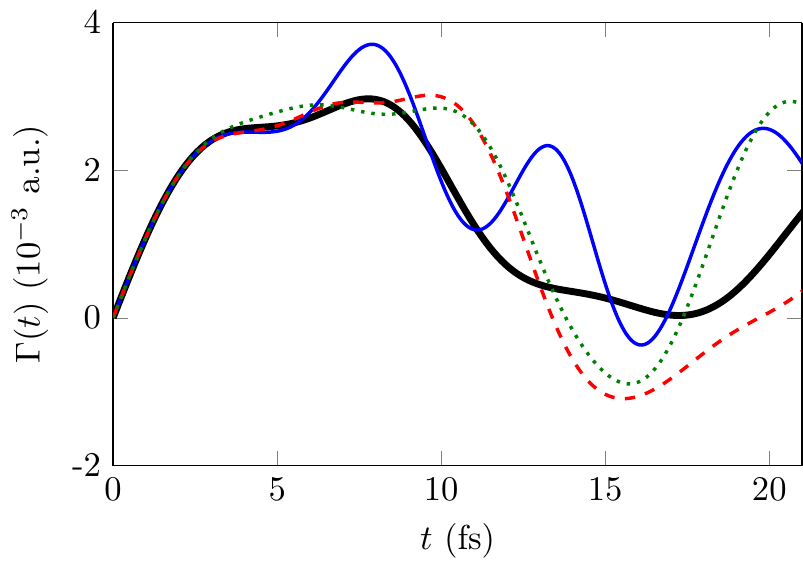}
	\caption{(Color online) Sum of the canonical decoherence rates (Eq.(\ref{Eq:Partial_gamma})) for the field-free (thick solid black line) and controlled system: thin solid blue for C1-1, dotted green for C2-2 and dashed red for C1-2.}
	\label{fig:gamma}	
\end{figure}

During the field-free evolution, the volume of accessible states illustrated in Fig.\ref{fig:vol} decreases very fast, in about 30 fs with a smooth monotonous decreasing profile. Nevertheless the decay is not exponential as it should be in a Markovian process. The duration of the control is fixed to 20 fs, i.e. less than the time for a complete decay of the volume. The resulting behaviors are displayed with the three control strategies C1-1, C2-2 and C1-2. Basically, after 5 fs, the decay is slightly faster than the field-free case and, more importantly, one observes some bumps, considered as clear signatures of non-Markovianity. 
Actually, the bumps arise at times close to 12 fs (for C1-1) or 17 fs (for C2-2) which could be associated with the maxima of the Stark shifts affecting the system energy gaps as displayed in Fig.\ref{fig:pop}.
As shown in Eq.(\ref{Eq:Vol-1}), this volume can also be computed from the sum of the canonical rates which are the eigenvalues of the decoherence matrix. This sum (Eq.(\ref{Eq:Vol-2})) displayed in Fig.\ref{fig:gamma}  clearly shows the increase of non-Markovianity during the controlled evolution.
More precisely, negative values for $\Gamma(t)$, responsible for the bumps of the volume, occur between 12 fs and 17 fs, mainly with the C2-2 and C1-2 control strategies. It is worthwhile noting the relation with important Stark shift affecting the system at such times as seen on Fig.\ref{fig:pop}.
\begin{figure}[!ht]
	\centering
	\includegraphics[width=0.95\textwidth]{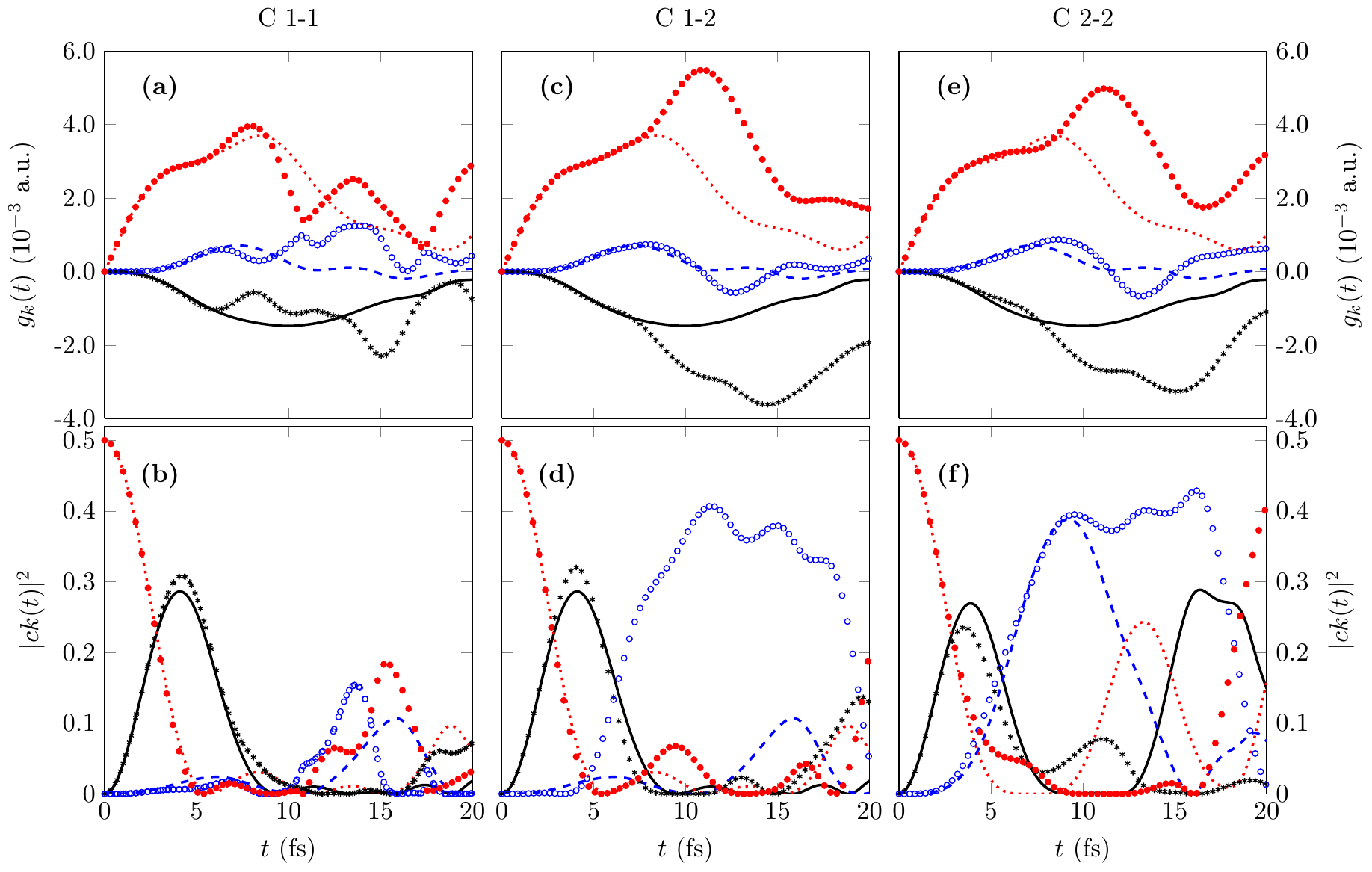}
	\caption{(Color online) Panel (a): canonical rates (Eq.(\ref{Eq:rho-3} ) for the field-free (solid black $k=1$, dashed blue $k=2$ and dotted red $k=3$ lines) and field-controlled (black stars, blue thin circles and red thick circles) evolution during the three control strategies. The rates are given in ascending order. Panel (b) : weights of the decoherence channels $ {c_k}(t) $  during the same evolution. }
	\label{fig:rates}
\end{figure}

\noindent
These analyses conclude that the field-dressed dynamics during the optimal control is more non-Markovian than the field free evolution. Moreover, one may question about the particular role of the quantum channel with the negative rate that should correspond to some backward flow. In order to observe the role of the different decoherence channels (Eq.(\ref{Eq:rho-3})) during the evolution of a given initial state, we compute the weight of the three quantum channels as: 
\begin{equation}\label{Eq:Channels}
{c_k}(t) = Tr\left[ {C_k^\dag (t)\rho (t)}\right].
\end{equation}

\begin{figure}[!ht]
	\centering
	\includegraphics{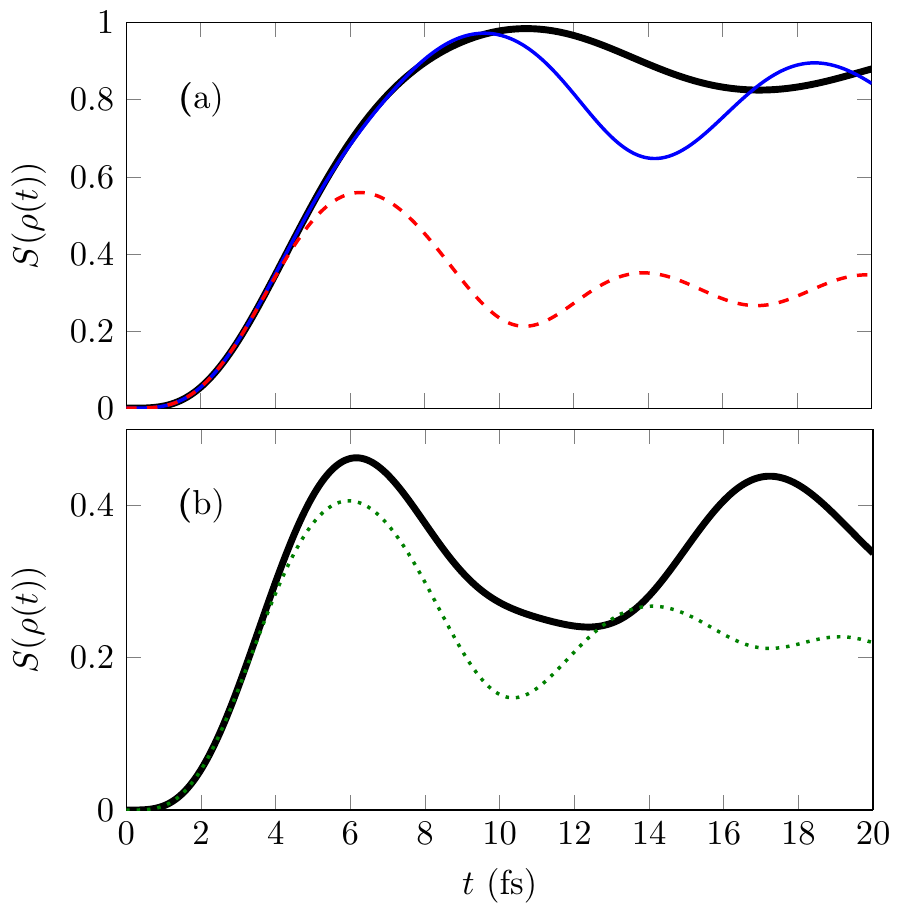}
	\caption{(Color online) System entropy during the field-free (solid lines) or field-controlled dynamics (dashed and dotted lines). Panel (a): initial state 1, where dashed blue dashed line is for the control strategy C1-1 and red dotted line is for C1-2. Panel (b) : Same as (a) but for the initial state 2 (control C2-2)}
	\label{fig:entropy}
\end{figure}
\noindent
Note that the operator $G_0$ (corresponding to the unity matrix) is not involved in the computation of coherence matrices so that the initial sum of $ |{c_k}(t=0) |^2$   is equal to 0.5 and this sum is not conserved during the evolution since the decoherence matrix only describes the volume decrease and not its translation in the Bloch sphere.  The upper panels of Fig.\ref{fig:rates} show the three canonical rates during the field-free and field-controlled evolution. The rates are given in increasing order so that channel $k=1$ corresponds to the negative rate, which may become even more negative during the control as seen in panels (c) and (e) after 7 fs during controls C1-2 and C2-2. The lower panels present the weights  $ |{c_k}(t) |^2$ during the relaxation. This illustrates the different impact of the negative rate during the control. The main observations are the following: (i) The weights of channel $k = 1$ with the most negative rates (black stars) always dominate around 5 fs but become the lowest after 8 fs except at the end of controls C1-1 and C2-2; (ii) The leading channel after 8 fs is $k=2$ associated with the smallest positive rates (blue curves) during C2-2 and C1-2. It decreases with respect to the field free case at the end of the C2-2 strategy; (iii) The highest positive rates are increased by the control fields, but more importantly their weights may decrease, for instance in the range 10-15 fs during control C2-2. As a consequence, the effective decay rate rate is basically affected by the combination of these effects. The increase of non-Markovianity during control does not necessarily imply that the channel with the negative rate plays the most significant role. In other words, an efficient control strategy for enforcing the bumps in the volume evolution, cannot merely be the tracking at each time of the channel with the negative partial rate, as it could be expected.

\noindent
The volume of reachable states is a global property of the system. It is built from the dynamical map so that when it exhibits non Markovian witness, it is expected that similar signatures could be found in properties related to the evolution of a particular initial state. As already discussed \cite{Haseli}, non-Markovian witness can be seen in the system entropy (Eq.(\ref{Eq:Entropy})) shown in Fig.\ref{fig:entropy}. The Markovian evolution of the entropy when the initial state is a pure state should be a monotonous evolution towards the value associated to the final Boltzmann mixture. In the present case, due to the energy gap, the final state is nearly the ground eigenstate so that the entropy profile should be a monotonous bell shape function. The non-Markovian signature is linked to any local decrease in the entropy which corresponds to a similar local bump in the purity $Tr(\rho^2(t))$ and therefore to an enhancement of the coherence. For instance, such a non-Markovian information back flow occurs between 11 and 18 fs in the field free evolution of state 1 and between 6 and 12 fs for state 2 (black curves). One observes that the dressed dynamics enhances this effect and more interestingly reduces the maximum entropy in a given time interval as during the control 1 to 2 (red dots in Fig.\ref{fig:entropy}).

\section{Conclusion}\label{sec:conclusion}
This work is devoted to a detailed analysis of external field control versus dissipation in non-Markovian strongly coupled open quantum systems. A heterojunction is taken as an illustrative example with its specific parameters and spectral density, building up a spin-boson type Hamiltonian. With respect to methodology, the originality relies on a complete implementation of an optimal control scheme, together with a fully converged HEOM treatment of the master equation describing the time evolution of the two-level sub-system density matrix beyond a perturbative regime. 

\noindent
We put the emphasis on control scenarios aiming at producing physically relevant processes within the two-level sub-system interacting with its environmental bath. The ultimate goal is to protect against decoherence, the sub-system (such as a qubit), the control taking advantage from memory effects to draw back some information content from the bath to the sub-system. As a first attempt, we consider two targets, namely, the revival of an initial state $\vert i\rangle$ ($i=1,2$) or a transition between the two states of the sub-system. The optimal control is precisely concerned with these goals through the populations of these states given in terms of the diagonal elements $\rho_{11}(t)$ and $\rho_{22}(t)$ of the sub-system density matrix. Once such control fields have been found, we address the consequences on the bath memory responses. Basically, we observe that non-Markovianity is increased during the optimally driven process. This is actually quantified through some typical signatures: time-dependent behavior of the volume of accessible states displaying bumps during its monotonic decay or the time-dependent behavior of the entropy exhibiting transitory decreases. At that point we have shown that a control aiming at the protection against decay of the sub-system characteristics provides, as a consequence, higher non-Markovian response of the bath. However, one of the main conclusions is that the mechanism does not necessarily increase the component on the quantum decay channel with the negative rate. We observe in most of the cases a decrease of the weight of the channel with the largest decay rate. Similar behaviors have been obtained for other targets such as the one inducing relaxation towards the ground system eigenstate. The control performances remain however rather modest. The main reasons are the strong system-bath coupling and the limited range of our flash field amplitudes, in relation with their experimental feasibility. To go beyond such limitations, we have to refer to ultra short and intense laser pulses. This requires the introduction of an additional constraint in the optimal control scheme to correct the time integrated pulse area that, following Maxwell equations should be zero \cite{Brabec, Dion, Bandrauk,Atabek}. Finally, even more realistic calculations should be conducted with ab-initio transition dipoles, resulting from quantum chemistry codes.

\noindent
As mid-term perspectives, future works should deal with exerting control directly on bath dynamics, in such a way to decrease decoherence of the sub-system, or in other words, achieve appropriate control of non-Markovianity to better protect sub-system characteristics. To that end, different strategies can be proposed: (i) Additional control of the environment through the introduction of a transition dipole among bath normal modes; (ii) Extraction of a collective mode from the bath so as to deal with a control involving an augmented active system, as has already been done in field-free heterojunction \cite{Mangaud} or in a SQUID model \cite{Koch}. We are actively pursuing research of these topics.

\section*{Acknowledgment}
	
We acknowledge support from the ANR-DFG COQS, under Grant No. \mbox{ANR-15-CE30-0023-01.} This work has been performed with the support of the Technische Universit\"at M\"unchen Institute for Advanced Study, funded by the German Excellence Initiative and the European Union Seventh Framework Programme under Grant Agreement No. 291763.
	
	\section*{References}

\end{document}